\documentclass[11pt,english,english]{article}
\usepackage[T1]{fontenc}
\usepackage[latin1]{inputenc}
\usepackage{babel}
\usepackage{graphics}

\makeatletter

\providecommand{\LyX}{L\kern-.1667em\lower.25em\hbox{Y}\kern-.125emX\@}

\usepackage[T1]{fontenc}
\usepackage[latin1]{inputenc}
\usepackage{babel}
\usepackage{setspace}

\begin{document}

\title{Production of Pentaquark States in pp Collisions within the Microcanonical
Ensemble}

\author{F.M.Liu\( ^{1,2} \)%
\thanks{Research fellow of Alexander von Humboldt Foundation
}, H. Stöcker\( ^{1} \), K. Werner\( ^{3} \)}

\maketitle
\( ^{1} \)Institut f\"{u}r Theoretische Physik, J.W.Goethe Universit\"{a}t,
Frankfurt am Main, Germany

\( ^{2} \)Institute of Particle Physics, Central China Normal University,
Wuhan, ChinaInstituite of Particle Physics

\( ^{3} \)Laboratoire SUBATECH, University of Nantes - IN2P3/CNRS
- Ecole des Mines de Nantes, Nantes, France

\begin{abstract}
The microcanonical statistical approach is applied to study the production
of pentaquark states in pp collisions. We predict the average multiplicity
and average transverse momentum of \( \Theta ^{+}(1540) \) and \( \Xi (1860) \)
and their antiparticles at different energies. 
\end{abstract}
Recently an exotic baryon \( \Theta ^{+}(1540) \) with the quantum
numbers of \( K^{+}n \) has been reported in several experiments\cite{Japen-theta,penta2,penta3,penta4,penta5}.
The \( \Theta ^{+}(1540) \) can not be a three quark state. Its minimal
quark content is \( (uudd\bar{s}) \), a \( q^{4}\bar{q} \) pentaquark
state. 

Pentaquark states have theoretically investigated since a long time
in the context of the constituent quark model\cite{5old,strottman}.
Some of these are expected to have charge and strangeness quantum
number combinations that can not be explained by three quark-quark
states. A variety of models has been employed to construct \( q^{4}\bar{q} \)
pentaquark states differently and predict differently masses and quantum
numbers. For example, the chiral soliton (Skyrme) model\cite{chiral}
predicts that the lightest member of the SU(3)-flavour (\( \overline{10}_{f},\, \frac{1}{2}^{+}) \)-let
has \( m_{\Theta }=1540\, \mathrm{MeV} \). The reported \( \Theta ^{+}(1540) \)
agrees with the prediction remarkably well. The other members of the
(\( \overline{10}_{f},\, \frac{1}{2}^{+}) \) antidecuplet are isospin-multiplets
of \( N \), \( \Sigma  \) and \( \Xi  \). In an uncorrelated quark
model\cite{strottman}, in which all quarks are in the ground state
of a mean field, the ground state of \( q^{4}\bar{q} \) has negative
parity. This is in striking difference to the chiral soliton model.
In Jaffe's model\cite{jaffe}, \( \Theta ^{+}(1540) \) is recognized
as a bound state of an antiquark with two highly correlated spin-zero
\( ud \) diquarks. Hence the lightest \( q^{4}\bar{q} \) state can
not be \( \Theta ^{+}(1540) \) but belongs to the \( N \) isospin-multiplets
with minimal quark content, i.e. \( uudd\bar{u} \). Other models
regard \( \Theta ^{+}(1540) \) as a member of the SU(3) flavour \( (27_{f}) \)-let.
Missing members of the multiplet are assigned to reported particles\cite{27lets}. 

The common members of above mentioned models are \( \Theta ^{+}(1540) \)
and the multiplets of \( \Xi  \), which can be \( \Xi ^{--}(ddss\bar{u}) \),
\( \Xi ^{-}(dssq\bar{q}) \) , \( \Xi ^{0}(ussq\bar{q}) \) or \( \Xi ^{+}(uuss\bar{d}) \),
where \( q\bar{q} \) is a hidden quark-antiquark pair \( u\bar{u} \)
or \( d\bar{d} \). Recently the NA49 collaboration \cite{penta6}
also presented the results of a search of \( \Xi ^{--}(1860) \) and
\( \Xi ^{0}(1860) \).

The estimation of \( \Theta ^{+}(1540) \) and \( \Xi (1860) \) yields
at different collisions energies independent of abovementioned models
will be helpful for the search for pentaquark states from proton-proton
collisions in the ongoing experiments in SPS and RHIC. Some work has
been done using the statistical hadronization approach within grandcanonical
and canonical ensembles. However the system is small in proton-proton
collisions, and a microcanonical ensemble should be justified.

The dynamical model NEXUS has also been used to estimate the yields
of \( \Theta ^{+}(1540) \) and \( \Xi (1860) \), via employing the
microcanonical ensemble to hadronize the remnants (formed by spectator
quarks from the collisions)\cite{marcus}. Here we use the microcanonical
ensemble to study the production of pentaquark states in pp collisions.
The microcanonical parameters are well studied already in previous
work via fitting the \( 4\pi  \) yields of charged pions, proton
and antiproton. We organise the paper as followed: first we explain
the model, how pentaquark states are produced in pp collisions, then
we check how reliable the microcanonical parameters are, then we present
our results, the yields and average transverse momentum of \( \Theta ^{+}(1540) \)
and \( \Xi (1860) \) and their antiparticles, and finally we compare
our results to some other theoretical work and discuss our predictions
on SPS and RHIC experiments.

Here we use the microcanonical ensemble to study the production of
pentaquark states in pp collisions. In the microcanonical ensemble,
we consider the final state of a proton-proton collision as a {}``cluster''
characterized by its volume \( V \) (the sum of individual proper
volumes), its energy \( E \) (the sum of all the cluster masses)
and the net flavour content \( Q=(N_{u}-N_{\bar{u}},N_{d}-N_{\bar{d}},N_{s}-N_{\bar{s}}) \),
decaying {}``statistically'' according to phase space. More precisely,
the probability of a cluster to hadronize into a configuration \( K=\{h_{1},p_{1};\ldots ;h_{n},p_{n}\} \)
of hadrons \( h_{i} \) with four momenta \( p_{i} \) is given by
the micro-canonical partition function \( \Omega (K) \),\begin{equation}
\label{phase-space}
\Omega (K)=\frac{V^{n}}{(2\pi \hbar )^{3n}}\, \prod _{i=1}^{n}g_{i}\, \prod _{\alpha \in \mathcal{S}}\, \frac{1}{n_{\alpha }!}\, \prod _{i=1}^{n}d^{3}p_{i}\, \delta (E-\Sigma \varepsilon _{i})\, \delta (\Sigma \vec{p}_{i})\, \delta _{Q,\Sigma q_{i}},
\end{equation}
 with \( \varepsilon _{i}=\sqrt{m_{i}^{2}+p_{i}^{2}} \) being the
energy, and \( \vec{p}_{i} \) the 3-momentum of particle \( i \).
\( n_{\alpha } \) is the number of hadrons of species \( \alpha  \),
and \( g_{i} \) is the degeneracy of particle \( i \). The term
\( \delta _{Q,\Sigma q_{i}} \) ensures flavour conservation and the
net flavour content \( Q=(4,2,0) \); \( q_{i} \) is the flavour
vector of hadron \( i \). The symbol \( \mathcal{S} \) represents
the set of hadron species considered: The ordinary \( \mathcal{S} \)
contains the pseudoscalar and vector mesons \( (\pi ,K,\eta ,\eta ',\rho ,K^{*},\omega ,\phi ) \)
and the lowest spin-\( \frac{1}{2} \) and spin-\( \frac{3}{2} \)
baryons \( (N,\Lambda ,\Sigma ,\Xi ,\Delta ,\Sigma ^{*},\Xi ^{*},\Omega ) \)
and the corresponding antibaryons. We generate randomly configurations
\( K \) according to the probability distribution \( \Omega (K) \).
For the details see ref. \cite{wer}.

We add the pentaquark states \( \Theta ^{+}(1540) \), \( \Xi (1860) \)
and their antiparticles into \( \mathcal{S} \). The \( \Theta ^{+} \)
has quark contents \( (uudd\bar{s}) \). The \( \Xi (1860) \) can
be \( \Xi ^{--}(ddss\bar{u}) \), \( \Xi ^{-}(dssq\bar{q}) \) , \( \Xi ^{0}(ussq\bar{q}) \)
or \( \Xi ^{+}(uuss\bar{d}) \). The spin of pentaquark states can
not be determined by experiments yet, and it is generally accepted
they are spin-\( \frac{1}{2} \) particles, and their degeneracy factor
\( g=2 \).

In our approach, because of the heavy masses of the pentaquark states,
the hadron configurations containing them appear very rarely. Therefore,
if the pentaquark states \( \Theta ^{+}(1540) \), \( \Xi (1860) \)
are spin-\( \frac{3}{2} \) particles, then their yields will be twice
as that of spin-\( \frac{1}{2} \) according to Eq. (\ref{phase-space}),
and their average transverse momenta will not be effected by their
spins. Our simulation has proven this point. In the following we report
the results assuming they are spin-\( \frac{1}{2} \) particles.

For \( \Xi ^{-}(dssq\bar{q}) \) and \( \Xi ^{0}(ussq\bar{q}) \)
, the \( q\bar{q} \) can be \( u\bar{u} \) or \( d\bar{d} \). Some
consider the two multiplets of \( \Xi  \) as three-quark \( q^{3} \)
states, \( \Xi ^{-}(dss) \) and \( \Xi ^{0}(uss) \). In the microcanonical
calculation, we do not need to distinguish between \( q^{3} \) and
\( q^{4}\bar{q} \), the two cases yields the same results when the
masses and degeneracy factors are the same, because the \( q\bar{q} \)
of the same flavour does not play any role in conserving flavours
or charge in the microcanonical statistical hadronization approach.

The microcanonical parameters \( (E,\, V) \) for pp collisions at
a given energy \( \sqrt{s}/\mathrm{GeV} \) are obtained by fitting
the \( 4\pi  \) multiplicities of the most copiously produced particles
(\( p,\, \bar{p},\, \pi ^{+},\, \pi ^{-} \)) \cite{sqm2003,micropp}:

\begin{eqnarray*}
E/\mathrm{GeV} & = & -3.8+3.76\mathrm{ln}\sqrt{s}+6.4/\sqrt{s}\\
V/\mathrm{fm}^{3} & = & -30.0376+14.93\mathrm{ln}\sqrt{s}-0.013\sqrt{s}.
\end{eqnarray*}
After add pentaquarks states into the hadron set, we still check the
\( 4\pi  \) multiplicities of \( p,\, \bar{p},\, \pi ^{+},\, \pi ^{-} \),
see fig. 1. Adding pentaquarks states into the hadron set does not
change the yields of light particles, except for antiproton. About
10\% more antiprotons are produced to compensate the net baryon numbers
carried by the pentaquark states.

\begin{figure}
{\centering \resizebox*{!}{0.5\textheight}{\includegraphics{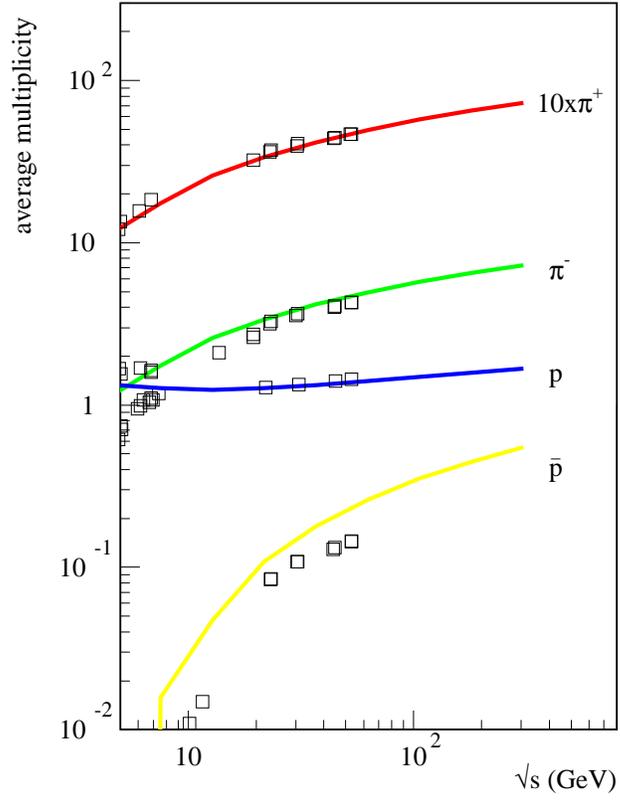}} \par}

\caption{\protect\( p,\, \bar{p},\, \pi ^{+},\, \pi ^{-}\protect \) excitation
functions. The empty square points are experimental data\cite{gia},
solid lines are microcanonical calculation after adding the pentaquark
states into the hadron set. }
\end{figure}

\begin{figure}
{\centering \resizebox*{0.9\textwidth}{!}{\includegraphics{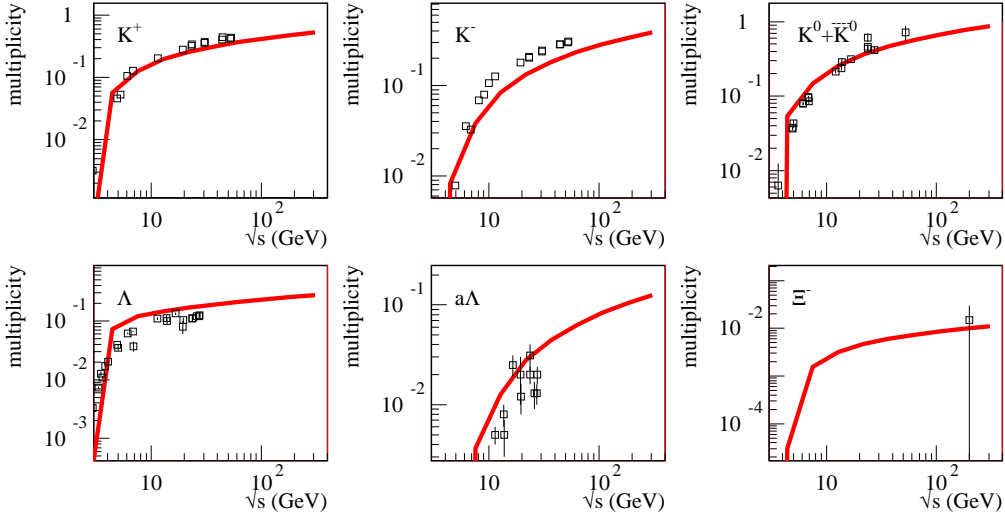}} \par}

\caption{\label{fig1}With a global factor \protect\( 1/3\protect \) to scale
strange hadrons such as K, \protect\( \Lambda \protect \) and \protect\( \bar{\Lambda }\protect \),
the microcanonical calculation(solid lines) can reproduce the data\cite{gia,gaz}(empty
squares). In the last plot, \protect\( \Xi ^{-}\protect \) from microcanonical
calculation scaled by factor \protect\( 1/9\protect \)(solid line)
agrees with data(empty square) \cite{ua5}. }
\end{figure}

Because the microcanonical calculation has no strangeness suppression
factor, so strange hadrons are overproduced\cite{micropp}. However,
with a global factor \( 1/3 \) to scale strange hadrons such as K,
\( \Lambda  \) and \( \bar{\Lambda } \), microcanonical calculations
(solid lines) are compared with the data\cite{gia,gaz} (empty squares),
c.f. Fig.  \ref{fig1}. Since \( \Xi  \) has two strange constituent
quarks, we should scale the yields with a factor \( 1/9 \). In Fig.
 \ref{fig1}, \( \Xi ^{-} \) from microcanonical calculation scaled
by factor \( 1/9 \) agrees with the UA5 data\cite{ua5}.

\begin{figure}
{\centering \resizebox*{0.5\textwidth}{!}{\includegraphics{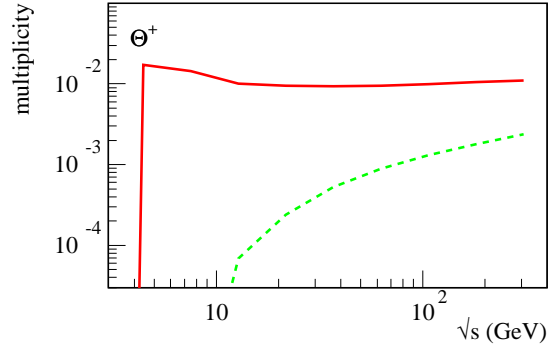}} \par}

\caption{\label{aa}The particle yields of \protect\( \Theta ^{+}\protect \)(solid
line) and its antiparticle (dashed line). }
\end{figure}

\begin{figure}
{\centering \resizebox*{0.9\textwidth}{!}{\includegraphics{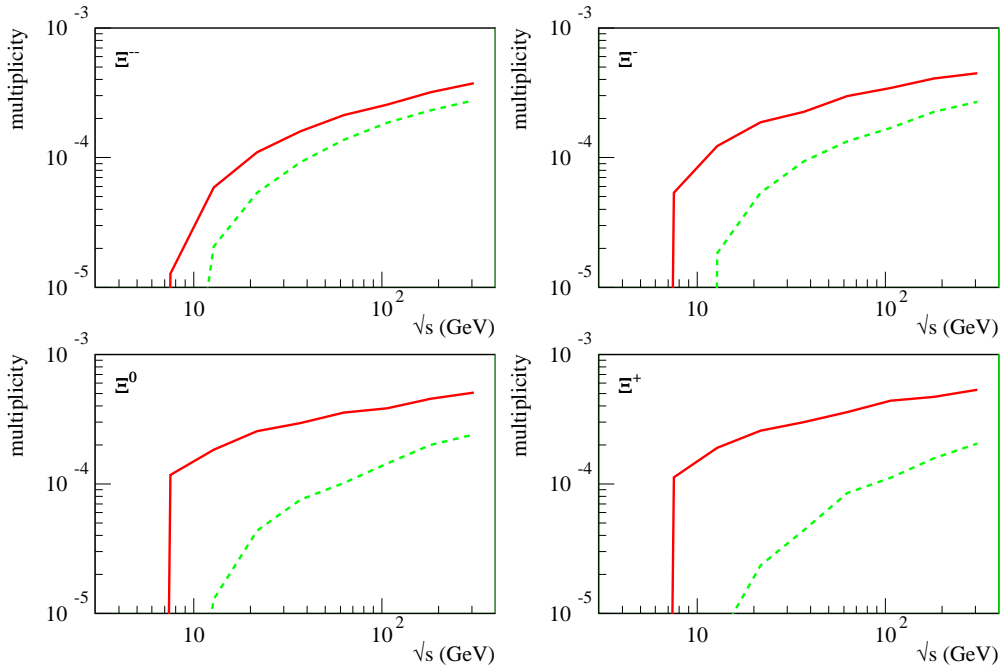}} \par}

\caption{\label{bb}The particle yields of \protect\( \Xi (1860)\protect \)
(solid lines) and their antiparticles(dashed lines).}
\end{figure}

The particle yields of \( \Theta ^{+} \)(solid line) and its antiparticle
(dashed line) from pp collisions at different collision energies are
shown in Fig. \ref{aa}. A factor \( 1/3 \) for \( \Theta ^{+} \)
and its antiparticle has been taken to account for the strangeness
suppression. The yields of the \( \Xi (1860) \) (solid lines) and
the antiparticles(dashed lines) are showed in Fig. \ref{bb}. A factor
\( 1/9 \) has been taken. With the increase of collision energy,
more and more pentaquarks states are produces, except \( \Theta ^{+} \).
We can see in Fig. \ref{aa}, \( \Theta ^{+} \)is favoured at low
energies because of the channel \( p+p\rightarrow \Theta ^{+}+\Sigma ^{+} \)
\cite{cs}.
 
\begin{figure}
{\centering \resizebox*{0.9\textwidth}{!}{\includegraphics{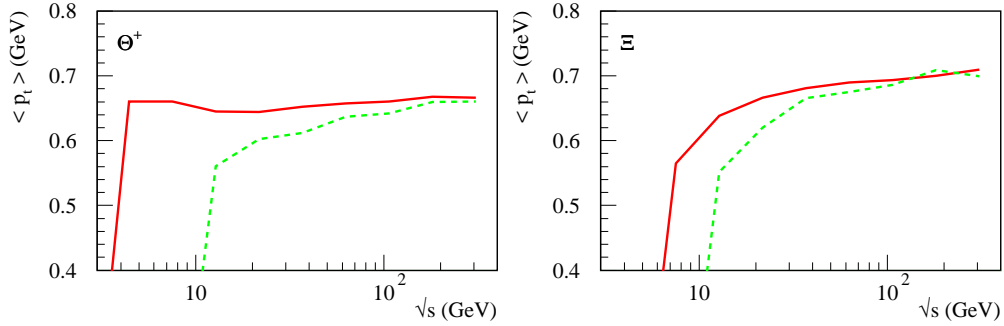}} \par}

\caption{\label{cc}The average transverse momentum of \protect\( \Theta ^{+}(1540)\protect \)
and \protect\( \Xi (1860)\protect \)(solid lines) and their antiparticles(dashed
lines).}
\end{figure}

With the \( 4\pi  \) yields of charged pions, protons and antiprotons
as input, the microcanonical calculation can predict reliably the
average transverse momentum of both non-strange and strange hadrons\cite{micropp}.
In Fig. \ref{cc} we also show the average transverse momentum of
\( \Theta ^{+}(1540) \) and \( \Xi (1860) \)(solid lines) and their
antiparticles(dashed lines). The difference between the average transverse
momentum of \( \Xi ^{--}(ddss\bar{u}) \), \( \Xi ^{-}(dss) \) ,
\( \Xi ^{0}(uss) \) and \( \Xi ^{+}(uuss\bar{d}) \) from microcanonical
calculation is ignorable. So do their antipartciles.
Now we compare our results to the previous research results at SPS
and RHIC energies. It is expected that the yields of pentaquark states
should be more than NEXUS\cite{marcus}. And we find indeed \( \Theta ^{+} \)
and \( \Xi (1860) \) are \( 2\sim 3 \) times more. The particle
ratio \( \Theta /p \) is about 0.7\%, which agrees suprisingly well
with the prediction from a quark molecular dynamics model prediction
0.6\%\cite{stefan}, while grandcanonical ensemble\cite{Randrup:2003fq}
estimation is about 6\%. The particle ratio \( \Xi ^{--}(1860)/\Xi ^{-} \)
is 2\% at SPS and 3\% at RHIC, which is \( 3\sim 4 \) bigger than
grandcanonical fitting \cite{Letessier:2003by}. 

The inclusive cross section \( \sigma _{pp\rightarrow \Theta ^{+}} \)
near the production threshold estimated with empirical coupling constants
and form factor is \( 20\, \mu b \)\cite{cs}, which is about ten
times smaller than our results. 

In conclusion, we presented a calculation of the yields of different
pentaquark states in pp collisions using the microcanonical approach.
We obtain roughly \( 10^{-2} \) (almost independent of energy) for
the \( \Theta ^{+} \), whereas the \( \Xi  \) yields increase strongly
with energy, reaching \( 4 \) x \( 10^{-4} \) at RHIC.

\noindent \textbf{Acknowledgements}

FML. Thanks Prof. B.Q. Ma and Prof. M. Bleicher for the fruitful discussions
and the Alexander von Humboldt Foundation for the finance support.

\end{document}